\documentclass[prd,twocolumn,10pt]{revtex4-1}
\usepackage{graphicx,color}
\usepackage{amsmath,amsfonts,amssymb}
\usepackage[]{hyperref}

\begin{document}

\title{The Nonlinear Field Space Theory}

\author{Jakub Mielczarek$^a$}\email{jakub.mielczarek@uj.edu.pl}
\author{Tomasz Trze\'{s}niewski$^{a,b}$}\email{tbwbt@ift.uni.wroc.pl}
\affiliation{${}^a$Institute of Physics, Jagiellonian University, {\L}ojasiewicza 11, 30-348 Krak\'{o}w, Poland \\
${}^b$Institute for Theoretical Physics, University of Wroc\l{}aw, pl.\ Borna 9, 50-204 Wroc\l{}aw, Poland}

\date{May 29, 2016}

\begin{abstract}
In recent years the idea that not only the configuration space of particles, i.e.\! spacetime, but also the corresponding momentum space may have nontrivial geometry has attracted significant attention, especially in the context of quantum gravity. The aim of this letter is to extend this concept to the domain of field theories, by introducing field spaces (i.e.\! phase spaces of field values) that are not affine spaces. After discussing the motivation and general aspects of our approach we present a detailed analysis of the prototype (quantum) Nonlinear Field Space Theory of a scalar field on the Minkowski background. We show that the nonlinear structure of a field space leads to numerous interesting predictions, including: non-locality, generalization of the uncertainty relations, algebra deformations, constraining of the maximal occupation number, shifting of the vacuum energy and renormalization of the charge and speed of propagation of field excitations. Furthermore, a compact field space is a natural way to implement the ``Principle of finiteness" of physical theories, which once motivated the Born-Infeld theory. Thus the presented framework has a variety of potential applications in the theories of fundamental interactions (e.g.\! quantum gravity), as well as in condensed matter physics (e.g.\! continuous spin chains), and can shed new light on the issue of divergences in quantum field theories.
\end{abstract}

\maketitle

\section{Introduction}
Depending on its type, the field theoretical description of Nature is assigning scalars, vectors, tensors 
or spinors to the points of space. The space of all possible values of a field, i.e.\! the field space, is a generalization of 
the particle phase space, with the number of degrees of freedom going to infinity. However, while nontrivial, 
curved phase spaces for particles and strings have been investigated in the context of quantum gravity \cite{Born:1938,Majid:2000,AmelinoCamelia:2011,Cianfrani:2014,Bojowald:2011} and string theory \cite{Tseytlin:1990,Freidel:2014,Freidel:2015}, 
the spaces of fields are typically assumed to be linear -- flat and infinite. The known exceptions are lattice field theories, defined on discretized 
spacetime \cite{Wilson:1974} and non-linear sigma models \cite{GellMann:1960,Witten:1984}, as well as their supersymmetric generalizations \cite{Zumino:1979}, where values of a multi-component scalar field (but usually not field velocities or momenta) are constrained to 
lie on a Riemannian manifold. 

In this letter we consider an extension of the standard field theory to the case when the whole field space is 
not a linear, affine space. By the field space we mean the space of values of the field and either field velocities 
in the Lagrange formulation, or field momenta in the canonical formulation. Here we will focus on the latter case. 

An important advantage of such a nontrivial structure of the field space is a possibility of restrictions on field values. This is 
encouraging since one can expect that for physical systems only finite values of fields are allowed, whereas in 
standard field theories arbitrary large values are possible, leading to different kinds of divergences. 

Thus we conjecture the standard Field Theory (FT) to be an approximation to the more general construction that ensures 
the finiteness of field values. The Nonlinear Field Space Theory (NFST) is a proposal for imposing the latter at the 
kinematical level. We do not rule out the consideration of NFST with unrestricted values of fields but here we will focus our attention on the compact case. 

Some steps towards imposing such a Principle of finiteness for field 
values were already made in the seminal paper \cite{Born:1934}, where M.~Born and 
L.~Infeld deformed the Lagrangian of electromagnetic field so that the field values became constrained, leading, e.g., to finite 
self-interaction energy of electron. However, in the Born-Infeld theory the field space is not compact but instead it is constrained dynamically by the special form of the Lagrange function. NFST is different and more similar 
to the case of a relativistic particle, where the maximal speed of propagation is a result of the spacetime geometry, 
independently of the particular form of the Lagrange or Hamilton function. For NFST it is the nonlinear structure of a given field 
space that determines the constraints on field values. 

Let us also stress that NFST should not be confused with Field Theories on Curved Spaces 
\cite{Birrell:1982} or the Group Field Theory \cite{Oriti:2012}. In the latter cases the field 
space is flat, while the background space(time), or momentum space, is curved. In NFST the field 
space may be curved but the background manifold is either flat or curved. In particular, in the 
example of NFST studied in the next section we assume that the background is Minkowski spacetime. 

\section{The scalar field}
To be more specific, let us consider the NFST for the simplest type of field -- the (real) scalar field. 
In the standard case the scalar field is a function $\phi: \mathcal{M} \rightarrow \mathcal{C}_{\phi} = \mathbb{R}$, where $\mathcal{M} 
= \Sigma \times \mathbb{R}$ is the spacetime manifold. Assuming $\Sigma = \mathbb{R}^3$, the field configuration space is 
$\mathcal{C} = \mathcal{C}_{\phi}^{4 \infty} = \mathbb{R}^{4 \infty}$. In the canonical formulation the field $\phi$ is accompanied by the canonical momentum 
$\pi: \mathcal{M} \rightarrow \mathbb{R}$, obeying the Poisson bracket $\left\{\phi({\bf x},t), \pi({\bf y},t)\right\} = \delta^{(3)}({\bf x - y})$. 
Then at every spacetime point the pair $(\phi,\pi)$ forms the phase space $\Gamma_{({\bf x},t)} = T^*(\mathcal{C}_{\phi({\bf x},t)}) = \mathbb{R} \times \mathbb{R}$ and the total phase space is given by $\prod_{({\bf x},t)} \Gamma_{({\bf x},t)}$. 

The idea of NFST is to generalize the phase space, so that $\Gamma_{({\bf x},t)} \neq \mathbb{R} 
\times \mathbb{R}$, which not necessarily has to be a cotangent bundle $T^*(\mathcal{C}_{\phi({\bf x},t)})$. 
At the kinematical level it is guaranteed (by virtue of the Darboux theorem) that the canonical 
symplectic form can be recovered on a sufficiently small neighbourhood in the phase space. However, in order to 
make a connection with the standard field theory we also have to define the dynamics in such a way that the 
proper form of the Hamiltonian is obtained in the limit of small values of $\phi$ and $\pi$. This limit 
can be defined with respect to e.g.\! the curvature scale inbuilt in a given NFST. As we discuss it in the Appendix (and it is similarly considered in \cite{Freidel:2014}), the notion 
of curved geometry of a phase space is indeed mathematically consistent, since a Riemannian metric can always be introduced there 
as an element of a so-called compatible triple \cite{Silva:2006}, although its explicit form is generally ambiguous. 

Alternatively, the NFST may be constructed for field variables defined in the Fourier space rather than position space, 
which actually turns out to be more convenient. To this end we perform the Fourier transform of the field:
\begin{equation}
\phi({\bf x},t) = \frac{1}{\sqrt{V}} \sum_{\bf k} \tilde{\phi}_{\bf k}(t) e^{i {\bf k \cdot x}}\,,
\end{equation}
and similarly for the momentum $\pi({\bf x},t)$, whose Fourier components will be denoted by $\tilde{\pi}_{\bf k}(t)$. 
In order to deal with the Kronecker rather than Dirac deltas we restricted the position space to a 3-volume $V$. 

The Fourier components are complex and, since the fields $\phi$ and $\pi$ are real, satisfy the so-called reality conditions. 
With the use of a suitable canonical transformation they can be, however, redefined so that we will work only with real variables. 
This can be achieved in many different ways but the most convenient transformation is given by:
\begin{equation}
\tilde{\phi}_{\bf k} = \frac{e^{i \frac{\pi}{4}} \phi_{\bf k} + e^{-i \frac{\pi}{4}} \phi_{-\bf k}}{\sqrt{2}}\,, \ \ 
\tilde{\pi}_{\bf k} = \frac{e^{i \frac{\pi}{4}} \pi_{\bf k} + e^{-i \frac{\pi}{4}} \pi_{-\bf k}}{\sqrt{2}}\,,
\end{equation}
where $\phi_{\bf k}, \pi_{\bf k} \in \mathbb{R}$ and $\left\{\phi_{\bf k}, \pi_{\bf k'}\right\} = \delta_{{\bf k}, {\bf k'}}$. 

Then, using the $\phi_{\bf k}$ and $\pi_{\bf k}$ variables, the standard Hamiltonian of a free massless scalar field:
\begin{equation}\label{Hclass}
H_{\phi} = \int_{V} d^3x \left( \frac{\pi^2}{2} + \frac{1}{2} \delta^{ab} \partial_a \phi\, \partial_b \phi \right)\,,
\end{equation}   
can be Fourier-transformed into
\begin{equation}\label{HclassF}
H_{\phi} = \frac{1}{2} \sum_{\bf k} \left( \pi_{\bf k}^2 + k^2 \phi_{\bf k}^2 \right)\,,
\end{equation}   
where $k = \sqrt{\bf k \cdot k}$. The Hamiltonian (\ref{HclassF}) is equivalent to an infinite sum of decoupled 
harmonic oscillators labelled by different wave numbers $k$. 

The field variables $\phi_{\bf k}$ and $\pi_{\bf k}$ span the phase space on which the 
Hamiltonian is defined. Since the modes are decoupled, the total phase space can be 
denoted by $\Gamma = \prod_{\bf k} \Gamma_{\bf k}$, where $\Gamma_{\bf k}$ is the 
phase space of a given mode. Usually the field configuration space $\mathcal{C}_{\bf k} 
= \mathbb{R} \ni \phi_{\bf k}$, and hence the phase space $\Gamma_{\bf k} = 
T^*(\mathcal{C}_{\bf k}) = \mathbb{R}^2 \ni (\phi_{\bf k}, \pi_{\bf k})$. Here, similarly to 
the case of position representation of $\phi$ and $\pi$, we will consider 
the situation when $\Gamma_{\bf k} \neq 
\mathbb{R}^2$.

\section{Spherical phase space}
In order to investigate the specific consequences of this framework let 
us present a concrete example and assume $\Gamma_{\bf k}$ to be a 2-sphere: $\Gamma_{\bf k} = S^2$. 
Such a phase space is indeed nontrivial: it cannot be decomposed into a product of two subspaces and is a 
compact manifold, which guarantees the finiteness of field variables that was discussed in the Introduction. Furthermore, 
it corresponds to the phase space of a spin (angular momentum), which allows us to use intuition from the 
atomic physics, and can also be compared with the fuzzy sphere geometry \cite{Madore:1992}.  

To define the Hamiltonian dynamics on the newly introduced phase space one first has to introduce a symplectic 
structure on it. Since $\text{dim}(\Gamma_{\bf k}) = 2$, the symplectic form $\omega$ can be naturally chosen 
as such that it is proportional to the area 2-form. By definition, this guarantees closure and non-degeneracy of 
the $\omega$ form, which is required for the definition of the Poisson bracket. 

Parametrizing the spherical phase space $\Gamma_{\bf k} = S^2$ by the standard angular variables $(\varphi,\theta)$ 
we then obtain the symplectic form $\omega = J \sin\theta\, d\varphi \wedge d\theta$, where $J$ is a free parameter 
of the dimension of angular momentum and $\int_{S^2} \omega = 4\pi J$, as expected for the area form. 
(The ``natural" metric on the symplectic manifold $(S^2,\omega)$ is introduced in the Appendix but 
the only role it plays in this letter is by giving the interpretation to $J$ as the inverse of the scalar curvature.) 
We assume that $J$ is ${\bf k}$-independent, which not necessarily has to hold in the general case. 

The compactness of phase space has profound consequences at the quantum level. Namely, 
since a single degree of freedom occupies the area of $2\pi \hslash$, the phase space having 
the area of $4\pi J$ can maximally accommodate $n_{\text{max}} := \frac{4\pi J}{2\pi \hslash} 
= \frac{2J}{\hslash}$ degrees of freedom. Consequently, the Hilbert space of 
such a system will be finite-dimensional.  

Before we pass to the more detailed analysis of the quantum theory let us relate the original 
phase space variables $(\phi_{\bf k},\pi_{\bf k})$ to the angular variables $(\varphi,\theta)$. In 
order to have the correct flat limit we choose
\begin{eqnarray}
(-\pi, \pi] \ni \varphi = \frac{\phi_{\bf k}}{R}\,, \ \ \text{and} \ \ [0, \pi] \ni \theta = \frac{\pi}{2} - \frac{R \pi_{\bf k}}{J}\,,
\end{eqnarray}
where $R$ is a constant introduced for dimensional reasons. With this redefinition the $\omega$ form rewrites to:
\begin{equation}
\omega = \cos \left( \frac{\pi_{\bf k} R}{J} \right) d\pi_{\bf k} \wedge d\phi_{\bf k}\,. 
\end{equation}
Clearly, for canonical momenta such that $\pi_{\bf k} \ll \frac{J}{R}$ the $\phi_{\bf k}$ and $\pi_{\bf k}$ variables become 
Darboux coordinates with the standard symplectic form $\omega = d\pi_{\bf k} \wedge d\phi_{\bf k}$. Furthermore, if we 
have a symplectic form the Poisson tensor $\mathcal{P}^{ij}$ can be defined as $\mathcal{P}^{ij} = (\omega^{-1})^{ij}$, 
allowing us to calculate the Poisson bracket $\{f,g\} = \mathcal{P}^{ij}
(\partial_i f)(\partial_j g)$. Hence the canonical Poisson bracket:
\begin{equation}\label{bracketqp}
\left\{\phi_{\bf k},\pi_{\bf k'}\right\} = \sec \left( \frac{\pi_{\bf k} R}{J} \right) \delta_{{\bf k},{\bf k'}}\,,
\end{equation}
which generalizes the standard one $\left\{\phi_{\bf k},\pi_{\bf k'}\right\} = \delta_{{\bf k},{\bf k'}}$. The bracket (\ref{bracketqp}) is, 
however, only locally well defined, because neither set of variables $(\phi_{\bf k},\pi_{\bf k})$ nor $(\varphi,\theta)$ is globally given 
on $S^2$ -- there is discontinuity at $\varphi = \pi$. Therefore, (\ref{bracketqp}) is not a good starting point for the 
quantization of the system. On the other hand, using $(\phi_{\bf k}, \pi_{\bf k})$ one can construct the 
well-known globally defined functions:
\begin{eqnarray}
J_x &:=& J \sin\theta \cos\varphi = J \cos \left( \frac{\pi_{\bf k} R}{J} \right) \cos \left( \frac{\phi_{\bf k}}{R} \right)\,, \label{Jx} \\
J_y &:=& J \sin\theta \sin\varphi = J \cos \left( \frac{\pi_{\bf k} R}{J} \right) \sin \left( \frac{\phi_{\bf k}}{R} \right)\,, \label{Jy} \\
J_z &:=& J \cos\theta = J \sin \left( \frac{\pi_{\bf k} R}{J} \right)\,, \label{Jz}
\end{eqnarray}
which form the $\mathfrak{su}(2)$ Lie algebra $\{J_i, J_j\} = \epsilon_{ijk} J^k$. 

Let us now discuss the kinematics of the quantized system. On the corresponding Hilbert space $\mathcal{H}_J$ (with a given value of $J$) 
we write the $\mathfrak{su}(2)$ algebra as $[\hat{J}_i, \hat{J}_j] = i \hslash \epsilon_{ijk} \hat{J}^k$. Then we have to take care of the issue of functional representations 
of states in $\mathcal{H}_J$, on which the operators $\hat{J}_i$ are acting. Due to the 
non-product form of the considered phase space, the field configuration and momentum representations of a quantum state 
will be meaningful only locally. Therefore, in general we should instead define a quantum quasiprobability distribution (which is not necessarily a positive definite function) 
on the phase space, such as the Wigner function. With the use of a Wigner function $W(\varphi,\theta)$ the expectation value of an operator $\hat{A}$ can be given as the phase space average $\langle \hat{A} \rangle 
:= \int_{S^2} d^2\Omega\, A(\varphi,\theta) W(\varphi,\theta)$. Following \cite{Stratonovich:1957,Agarwal:1981}, the Wigner function for a 
pure state $|\Psi \rangle \in \mathcal{H}_J$ on the spherical phase 
space can be defined as $W(\varphi,\theta) :=\text{tr}(|\Psi\rangle \langle\Psi| \hat{w}(\varphi,\theta))$,
where $\hat{w}(\varphi,\theta)$ denotes the Wigner operator. 

Analogously to the cylindrical phase space discussed in \cite{Bojowald:2011}, the $\hat{J}_i$ 
operators can be expanded in powers of the operators $\hat{\phi}_{\bf k}$ and $\hat{\pi}_{\bf k}$. 
Such a procedure is valid for phase space quasiprobability distributions 
(such as the Wigner function) supported on sufficiently small values of $\phi_{\bf k}$ and 
$\pi_{\bf k}$ ($\phi_{\bf k} \ll R \frac{\pi}{2}$, $\pi_{\bf k} \ll \frac{J}{R} \frac{\pi}{2}$). 
Locally, where the phase space can be approximately decomposed into a product of 
configuration and momentum spaces, this condition can be expressed in terms of supports 
of the configuration and momentum representations of a quantum state. In particular, the 
class of states which is expected to fulfill the conditions are the ``sufficiently peaked'' 
coherent states. Taking this into account, with the use of expressions (\ref{Jx}-\ref{Jz}), we find: 
\begin{eqnarray}
\hat{J}_x &=& J \left( 1 - \frac{1}{2R^2} \hat{\phi}_{\bf k}^2 - \frac{R^2}{2J^2} \hat{\pi}_{\bf k}^2 + \dots \right)\,, \\ 
\hat{J}_y &=& \frac{J}{R} \hat{\phi}_{\bf k} + \dots\,, \ \ \ \hat{J}_z = R \hat{\pi}_{\bf k} + \dots\,,
\end{eqnarray}
where dots denote higher powers of the $\hat{\phi}_{\bf k}$ and  $\hat{\pi}_{\bf k}$ operators. In the leading order, the commutator $[\hat{J}_y, \hat{J}_z] = i \hslash \hat{J}_x$ results in the following modified commutation relation:  
\begin{equation}\label{MCR}
[\hat{\phi}_{\bf k},\hat{\pi}_{\bf k}] = i \hslash \left( \hat{\mathbb{I}} - \frac{1}{2R^2} \hat{\phi}_{\bf k}^2 - \frac{R^2}{2J^2} \hat{\pi}_{\bf k}^2 + \dots \right)\,,
\end{equation}
where, due to the spectral theorem, for $O = \phi_{\bf k}, \pi_{\bf k}$ and $f(x) \in C^{\infty}$ the condition 
$\widehat{f(O)} = f(\hat{O)}$ is satisfied. 

One can, therefore, associate a nonlinear structure of the field phase space with a modification of the 
standard commutation relations. Furthermore, for the state in which $\langle \hat{\phi}_{\bf k} \rangle = 0 
= \langle \hat{\pi}_{\bf k} \rangle$ the commutation relation (\ref{MCR}) leads to the following generalized uncertainty principle:
\begin{equation}
\Delta \phi_{\bf k} \Delta \pi_{\bf k} \geq \frac{\hslash}{2} \left[1 - \frac{1}{2R^2} (\Delta \phi_{\bf k})^2 - \frac{R^2}{2J^2} (\Delta \pi_{\bf k})^2\right]\,.
\end{equation}
Inspection of this inequality reveals that (neglecting higher order corrections) due to the spherical field phase space either the $\Delta \phi_{\bf k}$ or $\Delta \pi_{\bf k}$ 
uncertainty can be saturated to zero while the other uncertainty is kept constant. A similar effect was observed in 
\cite{Bojowald:2011}, where the periodic phase space of the form $\Gamma = \mathbb{R} \times S^1$ was studied. 

\section{Dynamics}
Having the kinematics defined we are now ready to introduce the 
(classical) dynamics of the considered NFST. To this end we have to find a Hamiltonian which is satisfying two 
requirements: (i) it is a globally defined function on the phase space, (ii) it reduces to the Hamiltonian 
(\ref{HclassF}) in the flat phase space limit (i.e.\! for $J \rightarrow \infty$). 

In order to fulfill the condition (i) we can use globally defined variables $J_i$ as the Hamilton function's 
building blocks. Furthermore, the fact that in Nature one observes field excitations around $(\phi_{\bf k}, \pi_{\bf k}) = (0,0)$ 
suggests that this point in the phase space should be the classical minimum of the Hamiltonian. The goal 
of finding a Hamiltonian satisfying such a property together with the condition (ii) can be easily achieved by 
considering the formal analogy with a spin (magnetic moment) immersed in the constant magnetic field ${\bf B}$, 
which leads to a breakdown of the rotational invariance. Depending on the sign of the magnetic moment 
of a particle, the minimum energy state is associated with either parallel or anti-parallel alignment of the 
vectors ${\bf J} $ and ${\bf B}$. Consequently, we have $H \propto {\bf J \cdot B} = J_x B_x$, where the 
orientation of ${\bf B}$ has been chosen so that the condition (ii) is satisfied. Analogously, we define the 
Hamiltonian for our model in the following way:
\begin{eqnarray}
H_{\phi} &=& \sum_{\bf k} H_{\bf k}\,, \ \ \text{where} \\
\label{HS2}
H_{\bf k} &:=& -Jk \cos \left( \frac{\pi_{\bf k} }{\sqrt{Jk}} \right) \cos  \left( \sqrt{\frac{k}{J}} \phi_{\bf k} \right) \\
&=& -Jk + \frac{1}{2}\left( \pi_{\bf k}^2 + k^2\phi_{\bf k}^2 \right) - \frac{k}{4J} \phi_{\bf k}^2 \pi_{\bf k}^2 \nonumber \\
&-& \frac{1}{24Jk} \left( \pi_{\bf k}^4 + k^4\phi_{\bf k}^4 \right) + \mathcal{O}(J^{-2})\,, \nonumber
\end{eqnarray}   
where the condition (ii) is fixing $R = \sqrt{J/k}$ and $k$ in front of the cosines plays the formal role of $B_x$ from the spin example. 
In contrast to the classical case (\ref{HclassF}), the Hamiltonian (\ref{HS2}) is bounded both from below and above: $-Jk \leq H_{\bf k} \leq Jk$.   
However, its Taylor expansion shows that in the $J \rightarrow \infty$ limit the standard quadratic Hamiltonian is 
recovered up to the classically irrelevant constant contribution $-Jk$, which sets the lower energy bound. 

It is worth stressing that the Hamiltonian (\ref{HS2}) is, in some sense, similar to the one obtained by considering 
the so-called \emph{polymer quantization} \cite{Thiemann:2007} method, which arose from the Loop Quantum Gravity 
\cite{Ashtekar:2004} approach to quantum gravity. It has been shown in \cite{Hossain:2010} that in this 
quantization scheme the Hamiltonian for a given $k$-mode has the form $\hat{H}_{\bf k} = \frac{1}{2} 
\widehat{\frac{\sin^2(\lambda_k \pi_{\bf k})}{\lambda_k^2}} + \frac{1}{2} k^2 \hat{\phi}_{\bf k}^2$, where $\lambda_k$ 
is the $k$-dependent polimerization scale. The above expression has been so far considered only at the quantum level, 
where it is a consequence of the applied type of the Hilbert space -- the so-called Bohr space of almost periodic functions. 
However, there is a noticeable similarity of this result to the one that could be obtained for the $\Gamma_{\bf k} = \mathbb{R} \times S^1$ classical phase space. 

Finally, substituting the Hamiltonian (\ref{HS2}) to the Hamilton equations $\dot{f} = \left\{f, H_{\bf k}\right\}$, $f = \phi_{\bf k},\pi_{\bf k}$ we explicitly calculate that $\dot{\phi}_{\bf k} = {\sqrt{J k}}\tan \left( \frac{\pi_{\bf k}}{\sqrt{J k}} \right) \cos \left( \sqrt{\frac{k}{J}} \phi_{\bf k} \right)$ and $\dot{\pi}_{\bf k} = -\sqrt{J k}\, k \sin \left( \sqrt{\frac{k}{J}} \phi_{\bf k} \right)$. Due to the closure of the $\omega$ form one can also write $\omega = d \chi$, where $\chi = J \cos\theta\, d\varphi$ is a Liouville one-form, allowing us to perform the Legendre transformation of the Hamiltonian $H_{\bf k}$. Therefore the Lagrange formulation of the theory can be defined as well, with the Lagrangian $L_{\bf k} dt = {\sqrt{J k}} \sin \left( \frac{\pi_{\bf k}}{\sqrt{J k}} \right) d\phi_{\bf k} - H_{\bf k} dt$. 

The Hamiltonian in the position representation can be obtained by applying inverse Fourier transform to the expansion 
of (\ref{HS2}). Although a detailed discussion of this issue goes beyond the scope of this letter, a qualitative analysis allows 
us to observe that the extra interaction terms which appear in the position space representation of the Hamiltonian 
will have the non-local character. 

\section{Quantum dynamics}
It turns out that the Hamiltonian (\ref{HS2}) can be perturbatively diagonalized (at least up to the order $J^{-1}$) with the use of 
creation and annihilation operators. The procedure is almost completely analogous (similarly as in \cite{Hossain:2010}) to the case of standard interacting 
field theory, with only two differences. Firstly, the specific form of the interaction potential, which depends also on the field momentum. 
Secondly, due to the deformation in the commutation relation (\ref{MCR}), the expressions for the creation and annihilation 
operators $\hat{a}^\dagger_{\bf k}$, $\hat{a}_{\bf k}$ will differ from the usual ones. Furthermore, the $\hat{a}^\dagger_{\bf k}$ 
and $\hat{a}_{\bf k}$ representation of (\ref{MCR}) leads to the $q$-deformed version of their commutation relation: 
$\hat{a}_{\bf k} \hat{a}^\dagger_{\bf k} - q \hat{a}^\dagger_{\bf k} \hat{a}_{\bf k} = \hat{\mathbb{I}}$. Such a structure is directly 
related to the so-called generalized deformed oscillator algebras \cite{Bonatsos:1995}, which are connected with quantum groups. 

Namely, expressing the field operators as follows:
\begin{equation}\label{creanih}
\hat{\phi}_{\bf k} = \sqrt{\frac{\hslash}{2k}} \frac{\left( \hat{a}_{\bf k} + \hat{a}_{\bf k}^\dagger \right)}{\sqrt{1 + \frac{\hslash}{2J}}}\,, \ \ \hat{\pi}_{\bf k} = -i\sqrt{\frac{\hslash k}{2}} \frac{\left( \hat{a}_{\bf k} - \hat{a}_{\bf k}^\dagger \right)}{\sqrt{1 + \frac{\hslash}{2J}}}
\end{equation}
we calculate that the commutation relation (\ref{MCR}) introduces the $q$-deformation factor:
\begin{equation}
q = \frac{1 - \frac{\hslash}{2J}}{1 + \frac{\hslash}{2J}} = 1 - \frac{\hslash}{J} + \mathcal{O}(J^{-2})\,.
\end{equation}
The standard commutation relation of $\hat{a}^\dagger_{\bf k}$, $\hat{a}_{\bf k}$, as well as usual expressions for $\hat{\phi}_{\bf k}$ and $\hat{\pi}_{\bf k}$, are recovered in the $J \rightarrow \infty$ limit, as expected. 

The total Hilbert space of the system is $\mathcal{H} = \bigotimes_{\bf k} \mathcal{H}_{\bf k}$, where $\mathcal{H}_{\bf k} = \text{span} \left\{|0_{\bf k} \rangle, |1_{\bf k} \rangle, \dots, |n_{\text{max},\bf k} \rangle \right\}$. The actions of the $\hat{a}^\dagger_{\bf k}$ and $\hat{a}_{\bf k}$ operators on the $|n_{\bf k} \rangle$ basis states are found to have the form:
\begin{eqnarray}
\hat{a}_{\bf k}^\dagger |n \rangle = \sqrt{\frac{1 - q^{n+1}}{1 - q}} |n+1 \rangle\,, \ \ 
\hat{a}_{\bf k} |n\rangle = \sqrt{\frac{1 - q^n}{1 - q}} |n-1 \rangle\,, \nonumber
\end{eqnarray}
giving the $q$-deformed expression for the occupation number operator 
$\hat{a}^\dagger_{\bf k} \hat{a}_{\bf k} |n_{\bf k} \rangle = \frac{1 - q^n}{1 - q} |n_{\bf k} \rangle$. 

Then (\ref{creanih}) allows us to write down the perturbative expression for the quantum counterpart of the Hamiltonian (\ref{HS2}). Symmetrizing the $\phi_{\bf k}^2 \pi_{\bf k}^2$ term (which is equivalent to the choice of an operator ordering) we obtain:
\begin{eqnarray}\label{HquatnJ-1}
\hat{H}_{\bf k} &=& -Jk\, \hat{\mathbb{I}} + \left(\frac{1}{2} - \frac{\hslash}{4J} \right) k\hslash\, \hat{\mathbb{I}} + k\hslash \left( 1 - \frac{\hslash}{J} \right) \hat{a}^\dagger_{\bf k} \hat{a}_{\bf k} \nonumber \\   
&+& \frac{k\hslash }{24} \frac{\hslash}{J} \left( \hat{a}_{\bf k}^4 + (\hat{a}^\dagger_{\bf k})^4 - 6(\hat{a}^\dagger_{\bf k} \hat{a}_{\bf k})^2 - 6\hat{a}^\dagger_{\bf k} \hat{a}_{\bf k} - 6\hat{\mathbb{I}} \right) \nonumber \\
&+& \mathcal{O}(J^{-2})\,.
\end{eqnarray} 
The Hamiltonian can be decomposed into the free and potential part and, therefore, we can apply to it the time-independent perturbation theory. In the $1$-st order 
(the $J^{-1}$ contributions) this gives us the following eigenvalues $E_n^{(1)} = -Jk + k\hslash \left( n + \frac{1}{2} \right) - k\hslash \frac{\hslash}{4J} (1 + 3n + 3n^2) + \mathcal{O}(J^{-2})$ and the corresponding eigenstates $|n^{(1)} \rangle = |n \rangle - \frac{\hslash}{96 J} \sqrt{\frac{(n+4)!}{n!}} |n+4 \rangle + \frac{\hslash}{96 J} \sqrt{\frac{n!}{(n-4)!}} |n-4 \rangle + \mathcal{O}(J^{-2})$ for a given wave number $k$. 

We note that the vacuum energy in the new ground state 
$\langle 0_{\bf k}^{(1)}| \hat{H}_{\bf k} |0_{\bf k}^{(1)} \rangle = E_0^{(1)} = -Jk + \frac{1}{2}k \hslash - \frac{1}{4} k\hslash \frac{\hslash}{J} 
+ \mathcal{O}(J^{-2})$, is reduced with respect to the standard case not only by the factor $-Jk$ but also due to the contribution proportional to $J^{-1}$. Analysis of consequences of this effect in the context of the vacuum energy density and, presumably, cosmological 
constant problem is yet to be done. 

Another interesting observation is that while in the standard free  Quantum FT the field operator acting on the vacuum state is creating a single quantum $\hat{\phi}_{\bf k}(0) |0_{\bf k} \rangle = \sqrt{\frac{\hslash}{2k}} |1_{\bf k} \rangle$ with energy $E_1$, in the NFST the creation of a superposition of quanta can be naturally expected. Indeed, in the case considered here we have
\begin{equation}
\hat{\phi}_{\bf k}(0)\, |0_{\bf k}^{(1)} \rangle = \sum_n c_n |n_{\bf k}^{(1)} \rangle = c_1 |1_{\bf k}^{(1)} \rangle + c_3 |3_{\bf k}^{(1)} \rangle\,,
\end{equation}
where, up to the order $J^{-1}$, the coefficients  $c_1 = \sqrt{\frac{\hslash}{2k}} \left(1 - \frac{\hslash}{4J}\right)$ and 
$c_3 = -\sqrt{\frac{\hslash}{2k}}\, \frac{\hslash}{4\sqrt{6}J}$. 

Analysis of a two-point correlation function in the vacuum state $\mathcal{H} \ni |0 \rangle = \bigotimes_{\bf k} |0^{(1)}_{\bf k} \rangle$ provides us with the further interesting results. Namely, assuming statistical isotropy of the spatial field configurations, the two-point correlation function is given by
\begin{eqnarray}
\langle 0| \hat{\phi}({\bf x},t) \hat{\phi}({\bf y},t') |0 \rangle = \frac{1}{V} \sum_{{\bf k},n} |c_n|^2 e^{i {\bf k \cdot (x - y)} - i \Delta E_n (t - t')} \nonumber \\
= \frac{1}{V} \sum_{{\bf k}} \int \frac{d\omega}{2\pi} D_{(\omega,{\bf k})} e^{i {\bf k \cdot (x - y)} - i \omega (t - t')}\,, \nonumber
\end{eqnarray}
where (for a given wave number) $\Delta E_n = E^{(1)}_{n} - E^{(1)}_{0}$ and, denoting $p^2 = -\omega^2 + k^2$, we calculate the propagator: 
\begin{eqnarray}
D_{(\omega,{\bf k})} &=& \sum_{n} \frac{2i \Delta E_n |c_n|^2}{p^2 + \Delta E_n^2 - k^2 - i\epsilon} \nonumber \\
\label{prop}
&=& \frac{i \left( 1 - \frac{2}{J} \right)}{-\omega^2 + k^2 \left( 1 - \frac{3}{J} \right) + i\epsilon} + \mathcal{O}(J^{-2}) \\
&=& \frac{i}{-\omega^2 + k^2} + \frac{i}{J} \frac{k^2 + 2\omega^2}{(-\omega^2 + k^2)^2} + \mathcal{O}(J^{-2})\,, \nonumber
\end{eqnarray}
where for the purpose of transparency we set $c = 1$ and $\hslash = 1$. We note that the spherical field space leads to changes in the pole structure of the particle propagator. Consequently, the dispersion relation of field 
excitations associated with the propagator becomes modified. However, since 
$\Delta E_n$ naturally remains linear in $k$ at any order of the expansion in $J^{-1}$, the dispersion relation will always be linear as well. What is modified 
is the speed of propagation of an excitation. In particular, from the propagator given as the single term (\ref{prop}) one can deduce that the ``renormalized" 
speed of light reads $c_{\text{ren}} = 1 - \frac{3}{2} \frac{\hslash}{J} + \mathcal{O}(J^{-2})$. 

Finally, the propagator (\ref{prop}) can be used to predict the form of interaction potential between two point sources of 
the scalar field. Using the formula from \cite{Zee:2010} we find: 
\begin{eqnarray}
V(r) &=& 4 \pi i \int \frac{d^3k}{(2\pi \hslash)^3} e^{i {\bf k \cdot r}} D_{(0,{\bf k})} Q_0 \nonumber \\
&=& -\frac{Q_0}{r} \left( 1 + \frac{\hslash}{J} + \mathcal{O}(J^{-2}) \right)\,, 
\end{eqnarray}
where $Q_0$ is the charge of a field source. It is important to note that due to the ${\bf k}$-independence 
of J (scale invariance) the functional form of the $V(r)$ potential remains the same as in the standard case. 
The only difference is a ``renormalization" of the charge $Q_{\text{ren}} = Q_0 \left(1 + \frac{\hslash}{J} + \mathcal{O}(J^{-2})\right)$.

This effect, similarly to the renormalization of the speed of light discussed before, can be absorbed into the 
definition of variables, making the predictions possibly indistinguishable from the flat phase space results. 
Identification of the measurable quantities will be crucial in the context of physical applicability of NFST. 
Some of the potential empirical consequences of the proposed theory are discussed in the following section.

\section{Empirical consequences}
The general idea of nonlinear field spaces presented in this letter may turn out to be useful in different branches of theoretical physics, especially in the context 
of fundamental interactions and condensed matter. Depending on a 
particular system to which our framework is applied, the character of empirical 
predictions will differ. 

Let us stress once more that the construction of NFST does not affect spacetime itself but 
what is deformed instead is the phase space of field values. As the result, 
the corresponding field equations can exhibit different symmetries than the original 
spacetime, on which the theory has been defined. These symmetries reflect the effective structure of spacetime, as it is perceived by the field. Such a situation is well known in the condensed matter 
systems, where the effective symmetries are generally different from 
the background spacetime symmetries. 

Another important issue to note is that the prototype NFST discussed in Sections 3-5 concerns 
a single scalar field. While some of the features observed in this case can 
be expected to arise for other kinds of fields as well, extrapolations of the present results have to 
be taken carefully. Therefore, at the 
current stage it is premature to e.g. discuss the effects of our theory on the cross-sections for 
elementary particles but one can consider the Higgs or inflaton fields. 
 
The latter case indeed provides a promising testing arena for NFST, including the simple 
model discussed above. In order to apply it to the description of the generation of 
primordial perturbations, the Hamiltonian (\ref{HS2}) has to be generalized so that the 
cosmological evolution is taken into account. We find that for flat FRW cosmological background 
the Hamiltonian of a given mode has the following form: 
\begin{eqnarray}\label{Hcos}
H_{\bf k}^{\text{FRW}} &:=& -J a^3 \frac{k}{a} \cos \left( \sqrt{\frac{a}{Jk}}\, \frac{\pi_{\bf k}}{a^3} \right) \cos 
\left( \sqrt{\frac{k}{Ja}}\, \phi_{\bf k} \right) \nonumber \\ 
&=& -Jk a^2 + \frac{1}{2} \left( \frac{\pi_{\bf k}^2}{a^3} + 
a k^2 \phi_{\bf k}^2 \right) - \frac{k \phi_{\bf k}^2 \pi_{\bf k}^2}{4J a^4} \nonumber \\ 
&-& \frac{1}{24Jk} \left( \frac{\pi_{\bf k}^4}{a^8} + k^4 \phi_{\bf k}^4 \right) + \mathcal{O}(J^{-2})\,,
\nonumber 
\end{eqnarray}  
where $a$ denotes the scale factor. In the leading order this Hamiltonian leads to the same 
results as in the standard linear free field theory. However, due to the higher 
order interaction terms, a deviation from the Gaussian nature of cosmological 
perturbations is expected. Since interaction terms are even, the first nontrivial 
contribution to non-Gaussianity should appear at the level of connected four-point 
correlation function $\langle 0 | \phi_{\bf k_1} \phi_{\bf k_2} \phi_{\bf k_3} \phi_{\bf k_4} |0 \rangle_C \neq 0$, associated with the so-called trispectrum. 
The predicted non-Gaussianity (parametrized by $J$) can be the subject of observational 
constraints, e.g. with the data from the PLANCK mission \cite{Ade:2015}. 

Finally, in the model considered in this letter we assumed that the curvature scale of phase space $J$ is $k$-independent. 
It implies that the conformal invariance of Minkowski spacetime is preserved at 
the level of the field structure. This feature is reflected in expressions for 
the dispersion relation and interaction potential. In both cases, besides the 
renormalization of constants, the standard scalings corresponding to Minkowski 
spacetime are preserved. The conformal symmetry can be broken by introducing 
a $k$-dependence of the $J$ parameter (i.e.\! an additional scale). Then it 
is no longer expected that the dispersion relation and interaction potential preserve 
the standard scalings. The resulting deformed dispersion relation and the associated 
energy-dependence of the speed of propagation of field excitations could be 
constrained with the use of astrophysical observations \cite{AmelinoCamelia:1998}.

\section{Summary}
In recent years the idea that not only the configuration space but the whole phase space may have a nontrivial geometry has 
attracted significant attention, especially in different approaches to quantum gravity, where it can lead to testable predictions \cite{AmelinoCamelia:2011,AmelinoCamelia:2013}. 
The purpose of this letter was to extend this research direction into the domain of field theories, which 
for our use we call NFST. We have constructed a particular example of the NFST and showed that certain effects, which 
usually appear in the context of quantum gravity, emerge as a consequence of introducing nonlinearity of the field phase space. 

\section*{Acknowledgements}
This work is supported by the Iuventus Plus grant No.~0302/IP3/2015/73 from the Polish Ministry of 
Science and Higher Education. TT was additionally supported by the National Science Centre Poland, project 2014/13/B/ST2/04043.  

\section*{Appendix}
The classical phase space (of an isolated physical system) is a certain symplectic manifold $({\cal P},\omega)$: a manifold ${\cal P}$ equipped with a symplectic form $\omega$. It is known that on any $({\cal P},\omega)$ there exist almost complex structures, i.e.\! linear maps $I: T{\cal P} \rightarrow T{\cal P}$ satisfying the relation $I^2 = -1$. Furthermore, a symplectic manifold always possesses $\omega$-compatible almost complex structures, which means that for a given $I$ the map $g(.,.) \equiv \omega(.,I(.)): T{\cal P} \times T{\cal P} \rightarrow \mathbb{R}$ is a Riemannian metric on ${\cal P}$. $I$, $\omega$ and $g$ together are called a compatible triple \cite{Silva:2006}. 

In particular, for any oriented manifold ${\cal P}$ that can be embedded in $\mathbb{R}^3$ (like $S^2$) there exists the natural symplectic form and compatible almost complex structure, which are determined by the standard scalar and cross products on $\mathbb{R}^3$. Namely, for vectors $\vec{u}$, $\vec{v}$ tangent to ${\cal P}$ we have the expressions $\omega(\vec{u},\vec{v}) = \hat{n} \cdot (\vec{v} \times \vec{u})$, $I(\vec{u}) = \vec{u} \times \hat{n}$ (in the chosen sign convention), where $\hat{n}$ is a unit normal of ${\cal P}$. The resulting metric $g(\vec{u},\vec{v}) = \omega(\vec{u},I(\vec{v}))$ is the restriction to ${\cal P}$ of the standard Euclidean metric on $\mathbb{R}^3$ \cite{Silva:2006}. 

Let us apply this to the case of $S^2$ with the radius $\sqrt{J}$. Taking the vectors pointing in the directions of $\varphi$ and $\theta$, denoted by $\vec{u}_\varphi$, $\vec{u}_\theta$, we obtain $\omega(\vec{u}_\varphi,\vec{u}_\theta) = -\omega(\vec{u}_\theta,\vec{u}_\varphi) = J \sin\theta$, which gives $\omega = J \sin\theta\, d\varphi \wedge d\theta$ (as in Section 3). Similarly, we find that $I(\vec{u}_\varphi) = \sin\theta\, \vec{u}_\theta$, $I(\vec{u}_\theta) = -\sin^{-1}\theta\, \vec{u}_\varphi$ and hence in the matrix notation
\begin{align}
I = \left(
\begin{array}{cc}
0\! & -\sin^{-1}\theta \\ 
\sin\theta\! & 0
\end{array}
\right)\,.
\end{align}
The above $\omega$ and $I$ indeed determine the usual spherical metric $g = J (\sin^2\theta\, d\varphi^2 + d\theta^2)$, since $g(\vec{u}_\varphi,\vec{u}_\varphi) = J \sin^2\theta$, $g(\vec{u}_\theta,\vec{u}_\theta) = J$. However, in principle other compatible triples can be considered on $(S^2,\omega)$.

\end{document}